\documentclass[conference]{IEEEtran}
\IEEEoverridecommandlockouts
\usepackage{cite}
\usepackage{amssymb}
\usepackage{pifont}
\usepackage{algorithmic}
\usepackage{graphicx}
\usepackage{textcomp}
\usepackage{multirow}
\usepackage{xcolor}
\usepackage{bbding}
\usepackage{footnote}
\usepackage{amsmath}
\usepackage{tablefootnote}

\usepackage{hyperref}
\usepackage{float}
\usepackage{xcolor}
\usepackage{subcaption}
\hypersetup{
    colorlinks=true,
    citecolor=black,
    linkcolor=black,
    filecolor=black,      
    urlcolor=black,
}

\def\BibTeX{{\rm B\kern-.05em{\sc i\kern-.025em b}\kern-.08em
    T\kern-.1667em\lower.7ex\hbox{E}\kern-.125emX}}

\begin{document}

\title{Review-Incorporated Model-Agnostic Profile Injection Attacks on Recommender Systems}

\author{\IEEEauthorblockN{Shiyi Yang}
\IEEEauthorblockA{\textit{Data61, CSIRO and UNSW} \\
Sydney, Australia \\
shiyi.yang@data61.csiro.au}
\and
\IEEEauthorblockN{Lina Yao}
\IEEEauthorblockA{\textit{Data61, CSIRO, UNSW and} \\ 
\textit{Macquarie University}, Sydney, Australia \\
lina.yao@data61.csiro.au}
\and
\IEEEauthorblockN{Chen Wang, Xiwei Xu, Liming Zhu}
\IEEEauthorblockA{\textit{Data61, CSIRO} \\
Sydney, Australia \\
\{chen.wang, xiwei.xu, liming.zhu\}@data61.csiro.au}
\and
}

\maketitle

\begin{abstract}
Recent studies have shown that recommender systems (RSs) are highly vulnerable to data poisoning attacks. 
Understanding attack tactics helps improve the robustness of RSs. We intend to develop efficient attack methods that 
use limited resources to generate high-quality fake user profiles to achieve 1) transferability among black-box RSs 2)  and imperceptibility among detectors. 
In order to achieve these goals, we introduce 
textual reviews of products to enhance the generation quality of the profiles. Specifically, we propose a novel attack framework named R-Trojan, which formulates the attack objectives as an optimization problem and adopts a tailored transformer-based generative adversarial network (GAN) to solve it so that high-quality attack profiles can be produced. 
Comprehensive experiments on real-world datasets demonstrate that R-Trojan greatly outperforms state-of-the-art attack methods on various victim RSs under black-box settings and show its good imperceptibility.

\end{abstract}


\section{Introduction} \label{introduction}


In data poisoning attacks (aka shilling/profile injection attacks) to recommender systems (RSs) \cite{gunes2014shilling},
unscrupulous parties inject a group of fake user profiles including carefully-crafted user-item interaction data (e.g., ratings) to spoof a target RS for malicious purposes, such as promoting their own products or demoting competitors' products. 
Such attacks challenge the trustworthiness of RSs and threaten the fairness of the platform.
Hence, more efforts are being devoted to studying how RSs can be attacked to identify potential risks \cite{gunes2014shilling, li2016data, fang2018poisoning, christakopoulou2019adversarial, zhang2021attacking, lin2020attacking, huang2021data, wu2021triple, lin2022shilling}.




Many RSs have restrictions on accessing the details of their algorithms.
The architectures and parameters of the target RS are often not available to attackers in practice \cite{lin2022shilling}. 
Attackers often construct attack profiles that are difficult to distinguish from real profiles to increase the impact time and scope \cite{zhang2021attacking}. With limited resources to exploit, it is challenging to build high-quality attack profiles that have the following two properties: \textbf{transferability} (i.e., effective against different black-box RSs) and  \textbf{imperceptibility} (i.e., difficult to be detected).

It is difficult to transfer conventional shilling attacks (e.g., Random \cite{gunes2014shilling}) and some algorithm-specific poisoning attacks (e.g., PGA \cite{li2016data}) to deep learning (DL)-based RSs. On the other hand, many attack methods are poorly imperceptible, due to their optimization goals focus mainly on attack performance.
To address these problems, a few recent works (e.g., TrialAttack \cite{wu2021triple}  and Leg-UP \cite{lin2022shilling}) exploit generative adversarial networks (GANs) \cite{goodfellow2020generative} to produce effective fake user profiles. However, the existing attack profiles \cite{gunes2014shilling, li2016data, fang2018poisoning, christakopoulou2019adversarial, zhang2021attacking, lin2020attacking, huang2021data, lin2022shilling, wu2021triple} focus only on generating fake ratings to evaluate the robustness of RSs. 

Textual reviews are increasingly incorporated to boost recommendation performance \cite{xu2021understanding}, meanwhile, fake reviews with textual descriptions can easily influence user decisions and hence help businesses promote their products \cite{mohawesh2021fake}. How fake review text may affect RSs still remains an open problem.

We propose to use publicly available textual reviews to improve the generation quality of fake user profiles under limited available resources. 
Compared to numerical ratings, textual reviews contain rich semantic information \cite{zheng2017joint}, which better reveals user behaviors and properties of their preferred items than a numerical value using the Likert-scale ratings \cite{lin2022shilling}. In addition, the review text explains the reasons for the user’s rating, which is more convincing than mere ratings.


Specifically, we give a novel review-enhanced attack framework named R-Trojan to optimize attack objectives on transferability and imperceptibility. It uses a tailored transformer-improved GAN that consists of a generation module, a surrogate module and a detection module. The generation module builds high-quality fake user profiles, where fake user ratings are produced by reconstructing sorted real user profile templates. The corresponding fake textual reviews are generated by prompting a fine-tuned language model with given sentiments and topics. The surrogate module is a locally-trained review-based RS used to accurately measure the effectiveness of attacks with limited knowledge, for guiding the generation module to create transferable profiles. The detection module, on the other hand, is used to make the profiles imperceptible.


Our main contributions are summarized as follows:

\begin{itemize}
\item We introduce textual reviews to reinforce the generation quality of fake user profiles for attacking black-box RSs.
To the best of our knowledge, we are the first to attempt to introduce the reviews to guide profile generation.
\item We identify a new vulnerability by providing a pioneer study on poisoning attacks to review-based RSs.
\item We propose a novel attack framework R-Trojan to produce transferable and undetectable fake user profiles that contain fake numerical ratings and fake textual reviews.
\item Our experiment results on real-world datasets show that R-Trojan outperforms state-of-the-art attacks against various black-box RSs and has good detection escape ability.

\end{itemize}

\section{Related Work} \label{related_work}

\begin{figure*}[t]
    \centering
    \includegraphics[width=0.9\linewidth]{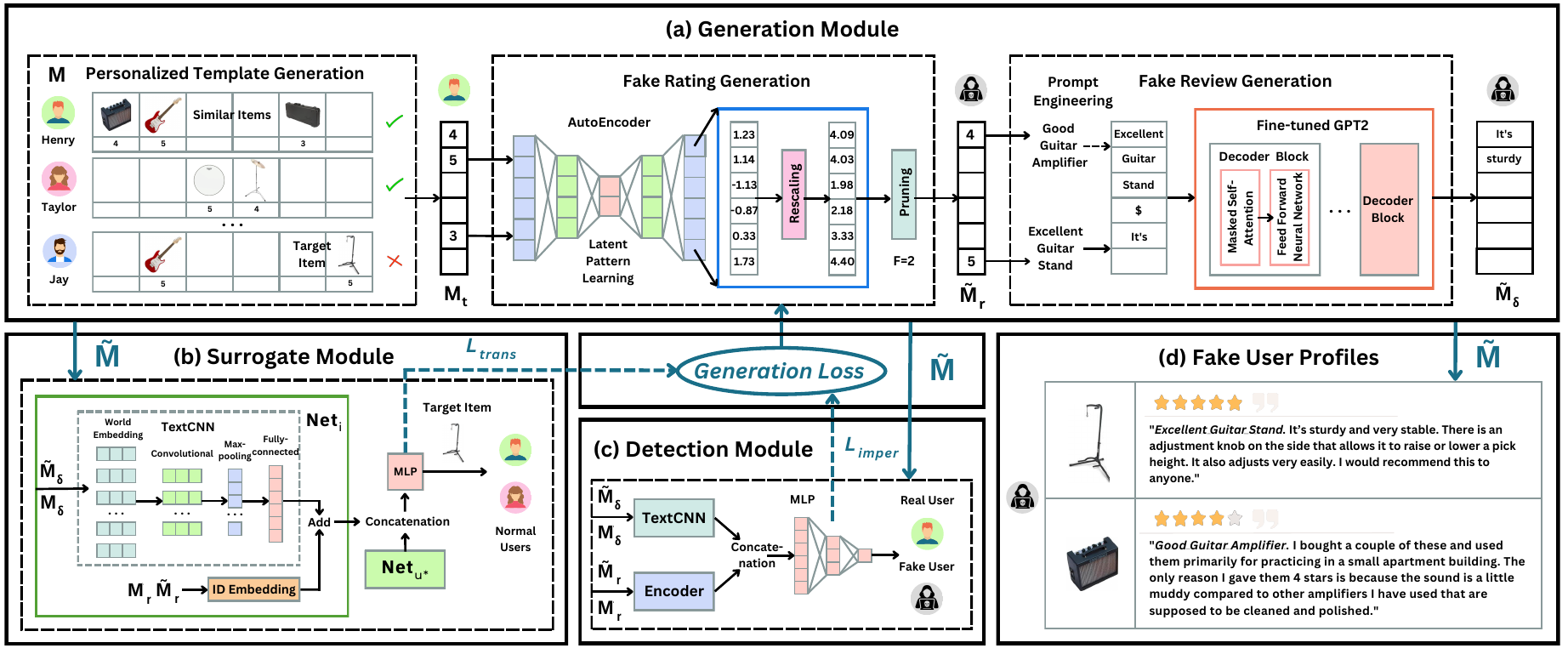}
    \vspace{-0.6em}
    \caption{R-Trojan Attack Framework}
    \label{fig:r_trojan}
    \vspace{-1.8em}
\end{figure*}

Conventional shilling attacks \cite{gunes2014shilling} such as Random and Bandwagon 
rely on global statistics and work mainly for traditional collaborative filtering (CF) (e.g., user-based KNN) \cite{lin2022shilling}. The attacks are simple heuristics-based and not transferable among different RSs (e.g., item-based KNN) \cite{lin2020attacking}. The lack of diversity in data generation makes them easy to be detected \cite{lin2022shilling}. 
Some algorithm-specific poisoning attacks are proposed to optimize for specific types of RSs,  such as matrix factorization-based \cite{li2016data} and graph-based \cite{fang2018poisoning}. However, their effectiveness to DL-based RSs \cite{zhang2019deep} is limited. An attack \cite{huang2021data} is proposed for DL-based under gray-box settings, which overestimates its practical performance. Since the methods are primarily concerned with optimizing attack performance, they lack of imperceptibility. Recently, more efforts have been made to leverage GANs for effective profile generation, e.g., DCGAN \cite{christakopoulou2019adversarial}, AUSH \cite{lin2020attacking}, TrialAttack \cite{wu2021triple}, RecUP \cite{zhang2021attacking} and Leg-UP \cite{lin2022shilling}. However, their attack profiles contain only ratings, which is not applicable to review-based RSs. To address these issues, we propose R-Trojan, a review-enhanced attack framework.

\section{Problem Formulation}

\subsection{Threat Model} \label{threat_model}

\subsubsection{\textbf{Attacker's Objective}} 
The first objective is \textit{Transferability}. There are two common types of shilling attacks used by unscrupulous parties for profits: push attacks and nuke attacks \cite{gunes2014shilling}. Considering that the two attacks are mutually convertible  \cite{huang2021data}, we focus on push attacks for simplicity. 
Let $h(t)$ to denote the \textit{hit ratio} of a target item $t$, which is the percentage of normal users whose top-$K$ recommendation lists include $t$ after the attack. Hence, our first goal is to maximize $h(t)$ on different black-box RSs. The second one (\textit{Imperceptibility}) is to make our attack as imperceptible as possible to maximize its impact time and thus expand the range of affected users.


\subsubsection{\textbf{Attacker's Knowledge}} We assume that the attacker has access to \textit{the training data $m\cdot n$ of victim RS}, where $m$ and $n$ are the number of users and items, respectively.
This is realistic as review data of RSs is often accessible by all users. 
However, the algorithms and models in real-world RSs are often complex and flexible. The implementation details and model parameters are not accessible\cite{lin2022shilling}. 
We hence treat the victim RS as a black-box.

\subsubsection{\textbf{Attacker's Capability}} \label{attacker_budget}

To avoid being detected while conducting attacks under a budget, an attacker often injects a limited number of profiles \cite{lin2022shilling} and interacts with a few items in each profile \cite{huang2021data}. We use $A$ to denote the attack size, which is the number of fake user profiles, and $F$ to denote the profile size, which is the number of interactive items in each profile.

\subsection{Formulate Attacks As An Optimization Problem}

We formulate the attack as a bi-level optimization problem \cite{lin2022shilling}. High-quality fake user profiles are obtained by solving this problem. Our attack consists of a generation, a surrogate (RS) and a detection (DE) module. \textit{The lower-level} computes the optimal parameters of the surrogate module (denoted by $\Theta$) and the detection module (denoted by $\Phi$) with the given normal matrix $M$ and fake matrix $\widetilde{M}$. 
An entry in $M$ is denoted by $(r_{ui}, \delta_{ui})$, in which $u$ is the row No. and $i$ is the column No. $r_{ui}$ is the numerical rating from user $u$ on item $i$ and $\delta_{ui}$ is the corresponding  review text. 
Each entry 
 of $\widetilde{M}$ is $(r_{\widetilde{u}i}, \delta_{\widetilde{u}i})$, where $r_{\widetilde{u}i}$  and $\delta_{\widetilde{u}i}$ are fake ratings and fake reviews produced by the generation module, respectively, while the entries of $\widetilde{M}$ in the existing attacks \cite{gunes2014shilling, li2016data, fang2018poisoning, christakopoulou2019adversarial, zhang2021attacking, lin2020attacking, huang2021data, lin2022shilling, wu2021triple} only contain $r_{\widetilde{u}i}$.
\textit{The upper-level} optimizes $\widetilde{M}$ to maximize the attack objectives based on model parameters obtained by solving the lower-level problem. The formulation is as below.

\vspace{-1.2em}
\begin{equation}
\begin{aligned}
\mathop{\min}_{\widetilde{M}}\ \lambda  L_{trans}(\widehat{M_{\Theta}}) + (1-\lambda) L_{imper}(\widehat{\widetilde{M}_{\Phi}}) \\
\text{subject to}\ \Theta = \mathop{\arg\min}_{\Theta} L_{RS}(M^*, \widehat{M_{\Theta}^*})\\
\text{and}\ \Phi = \mathop{\arg\max}_{\Phi} L_{DE}(M^*, \widehat{M_{\Phi}^*}),
\end{aligned}
\vspace{-0.9em}
\end{equation} where $M^*=\text{concatenation}(M;\widetilde{M})$, and  $\widehat{M_{\Theta}^*}$ and $\widehat{M_{\Phi}^*}$ are predictions from the corresponding models with parameters $\Theta$ and $\Phi$, respectively. $L_{RS}$ and $L_{DE}$ denote the training objectives of the related modules. $L_{trans}$ is the transferability objective defined on normal user's predictions $\widehat{M_{\Theta}}$, $L_{imper}$ is the imperceptibility objective defined on fake user's predictions $\widehat{\widetilde{M}_{\Phi}}$ and $\lambda$ is a configurable parameter to adjust the trade-off of two objectives when optimized simultaneously.



\section{R-Trojan}

In this section, we present a new attack framework R-Trojan, which is a three-module-based GAN, as shown in Fig. \ref{fig:r_trojan}.

\subsection{Generation Module}


\subsubsection{\textbf{Personalized Template Generation}} The sub-module $G_T$ aims to produce templates that contain real user patterns and are customized specifically for the target item $t$ for subsequent attack profile generation. Some works (e.g.,\cite{zhang2021attacking}) create attack profiles from scratch (e.g., noise). The resulting profiles can be of low quality without representing real user behaviors. To improve the imperceptibility of attack profiles, we sample a sub-matrix $M_t \in \mathbb{R}^{\widetilde{m} \times n}, \widetilde{m} = A$  from $M$ as templates, where $M_t$ contains only rating information. Some works (e.g., \cite{lin2022shilling}, \cite{lin2020attacking}) randomly sample templates. However, normal users generally do not interact with items randomly. As can be seen from the example in Fig. \ref{fig:r_trojan}, Henry is a guitarist who always buys guitar-related items, while Taylor is a drummer who doesn't buy guitar-related items. Therefore, using Talyor's profile as a template might be suspicious when $t$ is a `guitar stand'. To address this problem, we propose a strategy for creating personalized templates for $t$. Let $I_u = \{i \in n: r_{ui} \neq 0 \}$ denote the set of items that have been interacted by user $u$ and $M'$ is a copy of $M$ to facilitate operations. The steps of the strategy are as follows.

\begin{enumerate}
    \item Since the target normal users are users who have not yet interacted with $t$, R-Trojan \textit{removes} $u$'s profile from $M'$ if $t \in I_{u}$. As the example shows, Jay's profile will not be considered a template due to its interaction with $t$. 
    \item To comprehensively learn real user behaviors and increase the diversity of attack profiles, R-Trojan \textit{sorts} $M'$ based on $|I_u|$. As a result, profiles from the users with the largest interactions will be used as templates first.
    \item R-Trojan then leverages \textit{ attribute information (e.g., category $I_c$)} of the target item to obtain a \textit{similar item} set $I_s$, while $I_s = \{i \in n: i \in I_c \}$. $M'$ is \textit{sorted} by $I_s$ so that the profiles of users who have previously interacted with similar items will be prioritized. As the example shows, since $t$ is highly correlated to Henry's profile, R-Trojan uses such a template to improve attack imperceptibility. In addition, such users may also be the most vulnerable target group due to they have similar interests to $t$. As such, building fake profiles based on them can further improve the attack performance. Consequently, $M_t = G_T(M) = \{I_u \in M': u \leqslant A \}$.
\end{enumerate}

\subsubsection{\textbf{Fake Rating Generation}} The sub-module $G_R$ aims to generate effective fake ratings based on templates. To ensure the quality of attack profiles, we design three components.

\textit{\textbf{Pattern Learning.}} R-Trojan uses neural networks to capture  complex user-item associations from templates in the latent space. There are various possible paradigms for pattern learning. We use AutoEncoder \cite{zhang2019deep} with an encoder and a decoder module to implement this. 
As can be seen from Fig. \ref{fig:r_trojan}, the Encoder $E_{\psi_1}$ with parameters $\psi_1$ learns a low-dimensional feature representation (i.e., user preferences) of the input, while the Decoder $D_{\psi_2}$ with parameters $\psi_2$ regenerates the rating vectors from the representation. We use multi-layer perceptron (MLP) to build the Encoder and the Decoder:

\vspace{-1.9em}

\begin{gather}
   E_{\psi_1}(M_t) = f_{l_1}(\cdots f_2(f_1(M_t))\cdots), \\
   D_{\psi_2}(E_{\psi_1}(M_t)) = f_{2l_1}(\cdots f_{l_1+1}(E_{\psi_1}(M_t))\cdots),
\end{gather} where $f_j(\cdot), \ j=1, 2, ..., 2l_1-1$ denotes the mapping function for the $j$-th hidden layer. $f_j(x)=\text{ReLU}(W_jx+b_j)$, where ReLU is a non-linear activation function (i.e., $\text{ReLU}(z) = \text{max}(0, z)$), $W_j$ and $b_j$ are a learnable weight matrix and a bias vector for layer $j$ respectively. For the Encoder, we set the sizes of layers (that is, the dimension of $x$) to half the size of previous layers, but for the Decoder, we set the sizes of layers to twice the size of the previous layers. Moreover, batch normalization is added before the ReLU to decrease the generalization errors. Furthermore, a dropout strategy is introduced after the ReLU to prevent overfitting.

\textit{\textbf{Rescaling Output.}} To be consistent with the Likert-scale rating method \cite{lin2022shilling} commonly used in RSs, we design a function to rescale the output within the range of the rating (e.g., [1, 5]) for $f_{2l_1}(\cdot)$, as formulated below.
\vspace{-1.3em}

\begin{equation}
     \text{ReScale}(x) = \frac{\text{max}(M_r) - \text{min}(M_r)}{1+e^{-x}} + \text{min}(M_r),
     \vspace{-0.3em}
\end{equation} 
where $M_r \in \mathbb{R}^{m \times n}$  be a rating matrix of $M$. As a result, $f_{2l_1}(f_{2l_1-1}(\cdot)) = \text{ReScale}(W_{2l_1}f_{2l_1-1}(\cdot)$ $+b_{2l_1})$, where $W_{2l_1}$ and $b_{2l_1}$ are learnable parameters. As the example in the figure illustrates, the ratings are scaled up to a range of 1 to 5.

\textit{\textbf{Profile Pruning.}}
As discussed in Section \ref{threat_model}, the profiles of R-Trojan aim to maximize $h(t)$ and should contain only a few interactive items to  avoid detection. To this end, R-Trojan introduces a pruning strategy to improve the quality of profiles. If the item is $t$, the strategy gives it the maximum rating for item promotion. Otherwise, the strategy first executes the rounding-off operation $f_o(\cdot)$ to discretize the ratings to match the input characteristics of the RS and then performs masking operation $f_m(\cdot)$ to mask a large number of ratings to simulate the sparsity of the real data while maintaining actual patterns.

\vspace{-1.4em}

\begin{equation}\label{eq3}	
	\begin{aligned} 
 \text{Pruning}(d_{\widetilde{u}i}) = 
		&\left\{\begin{aligned}	
		&  \  \text{max}(M_r), \ \ \ \ \ \ \ \ \ \ \ \ \ \ \ \ \ \ \  \ \  \text{if} \ i = t; \\
		&  \ f_o(d_{\widetilde{u}i}) \odot f_m(|d_{\widetilde{u}i} - e_{\widetilde{u}i}|), \ \ \text{if} \  i \neq  t, \\
	\end{aligned}\right.
	\end{aligned}
 \vspace{-0.8em}
\end{equation} where $(d_{\widetilde{u}i})_{\widetilde{u} \in \widetilde{m}, i \in n}=D_{\psi_2}(E_{\psi_1}(M_t))$ and $(e_{\widetilde{u}i})_{\widetilde{u} \in \widetilde{m}, i \in n}=M_t$. $f_m(\cdot)$ indicates whether the item $i$ is to be retained ($f_m(\cdot)=1$) or not ($f_m(\cdot)=0$) as a filler item for the fake user $\widetilde{u}$. When the difference between $d_{\widetilde{u}i}$ and $e_{\widetilde{u}i}$ belongs to the first $F-1$ smallest differences and $e_{\widetilde{u}i} \neq 0$, $f_m(\cdot)=1$, otherwise $f_m(\cdot)=0$. As the example in Fig. \ref{fig:r_trojan} shows, when $F=2$, in addition to the target item, the first item is selected as a filler item for a profile, due to $|4.09-4|<|3.33-3|<|4.03-5|$. By doing so, the patterns of real users are preserved to the maximum extent, which further increases diversity (e.g., picky users) and thus the imperceptibility of profiles. To sum up, $\widetilde{M_r}  = G_R(G_T(M))$, $\widetilde{M_r}  \in \mathbb{R}^{\widetilde{m} \times n}$ is the rating matrix of $\widetilde{M}$.

\subsubsection{\textbf{Fake Review Generation}} The sub-module $G_V$ aims to generate human-like and insightful fake textual reviews for $\widetilde{M}$, by prompting a fine-tuned language model with given sentiments and topics, as shown in Fig. \ref{fig:r_trojan}(a).

\textit{\textbf{Prompting Engineering.}} Intuitively, word embeddings of real reviews can be fed into a neural network (e.g., AutoEncoder) to directly output word embeddings for fake reviews. However, a sentence that is composed of words mapped according to such embeddings in the vector space may not be coherent, which is suspicious and easy to be discovered. To avoid this problem, R-Trojan uses \textit{prompt engineering} to guide a large language model (LLM) \cite{radford2019language} towards generating desired responses. To make the generated fake reviews correspond to fake ratings to improve the imperceptibility of the attack, we craft prompts $\widetilde{P} = \{ \widetilde{u} \in \widetilde{m}, i \in n: p_{\widetilde{u}i} \}$ to control for the generated reviews $\delta_{\widetilde{u}i}$  containing sentence-level attributes (e.g., topics and sentiments), as formulated below. 

\vspace{-1em}
\begin{equation}
p_{\widetilde{u}i}=\text{concatenation}(f_s(r_{\widetilde{u}i});a_{i}),
\vspace{-0.4em}
\end{equation} where $f_s(r_{\widetilde{u}i})$ denotes rating information for sentiment attributes and $a_{i}$ presents product information for topic attributes, as shown in the figure. $f_s$ is the customer satisfaction  corresponding to Likert-scale, e.g, for 5 Point Likert-scale, if $r_{\widetilde{u}i}=1$, $f_s(\cdot)$ is `Very Poor'; if $r_{\widetilde{u}i}=2$, $f_s(\cdot)$ is `Poor'; if $r_{\widetilde{u}i}=3$, $f_s(\cdot)$ is `Average'; if $r_{\widetilde{u}i}=4$, $f_s(\cdot)$ is `Good'; if $r_{\widetilde{u}i}=5$, $f_s(\cdot)$ is `Excellent'. $a_i$ is the product name from set $I_a = \{ i \in n : a_{i} \}$.

\textbf{\textit{Fine-tuning LLMs.}} 
\textit{Fine-tuning} is a relatively cost-effective way for adapting pre-trained LLMs on proprietary data \cite{radford2019language}. Fine-tuning can achieve good performance on various downstream NLP tasks (e.g., text generation). As such, R-Trojan fine-tunes the parameters $\psi_3$ of a pre-trained autoregressive LLM based on publicly available and real-world review data. To further enhance the control of the sentence-level attributes, we add prompts $P = \{ u \in m, i \in n: p_{ui} \}$ as prefixes of the real reviews $\delta_{ui}$ so that the training set is conditional on the prompts. The formulation of $\psi_3$ obtained by minimizing the negative log-likelihood is given below.

\vspace{-1.5em}
\begin{equation}
     \psi_3 = \mathop{\arg\min}_{\psi_3} \ -\sum_{u \in m, i \in n} \sum_{k=1}^{|x_{ui}|} \log{p_{\psi_3}(x_{{ui}_k}|h_{{ui}_{<k}})},
\vspace{-0.5em}
\end{equation} where the activation at time step $k$ is $h_{{ui}_k}\in \mathbb{R}^{d_1}$, $h_{{ui}_k} = (h_{{ui}_k}^{(1)}; \cdots; h_{{ui}_k}^{(l)})$, which is a concatenation of all activation
layers at time step $k$, the last layer of $h_{{ui}_k}$ is used to compute the distribution for the next token: $p_{\psi_3}(x_{{ui}_{k+1}}|h_{{ui}_{\leqslant k}}) = \text{softmax}(W_{\psi_3}h_{{ui}_k}^{(l)})$, where $W_{\psi_3}$ is a trainable matrix that map $h_{{ui}_k}^{(l)}$ to logits over the vocabulary, and $x_{ui} = \text{concatenation}(p_{ui}; \delta_{ui})$.

Since the autoregression LLMs trained with a causal language modeling objective are more suitable for text generation tasks, R-Trojan uses GPT models to generate fake reviews. Among them, GPT-2 \cite{radford2019language} produces good sentence generation performance for our task and hence selected by R-Trojan, as shown in the figure.
As a result, $\widetilde{M_\delta} = \text{LLM}_\psi(\widetilde{P}) = G_V(G_R(G_T(M)))$, which is the corresponding review matrix, and it is combined with $\widetilde{M_r}$ to form $\widetilde{M}$.

\subsection{Surrogate Module}

Since victim RSs are generally black-boxes, a locally-trained surrogate RS is introduced to measure how effective the attack is, for improving attack transferability, as inspired by \cite{lin2022shilling}. To achieve more accurate measurement with limited knowledge, informative textual reviews are introduced. The module is a review-based RS called DeepCoNN++, which is based on a classical review model DeepCoNN \cite{zheng2017joint} but with some improvements, as shown in Fig. \ref{fig:r_trojan}(b). The model consists of two parallel neural networks, one for user modeling ($Net_{u^*}$, $u^* \in m \cup \widetilde{m}$), and another for item modeling ($Net_i, i \in n$). On the top of the two networks, a prediction model (i.e., MLP) is used to combine latent representations of users and items.


\textbf{\textit{Feature Modeling.}} We use TextCNN \cite{xu2021understanding} to extract user behaviors and item properties from user/item textual review documents, which is the same as DeepCoNN's text processing approach, as technically detailed in \cite{zheng2017joint}. 
DeepCoNN only uses review-based features, but not ratings-based features (i.e., ID embeddings) that are useful when there are few reviews. To enrich latent factors of users and items, we adopt both review-based and rating-based features, as formulated below. 

\vspace{-1em}
\begin{equation}
o_i =\text{TextCNN}_{\theta_1}(\delta_i) + q_i,
\vspace{-0.5em}
\end{equation} where $\delta_i$ is $i$'s review document that contains all reviews written for the item from both normal and malicious users,  $\text{TextCNN}_{\theta_1}(\delta_i) \in \mathbb{R}^{d_2\times1}$ denotes the review-based features, $q_i \in \mathbb{R}^{d_2\times1}$ is rating-based features and $o_i  \in \mathbb{R}^{d_2\times1}$ is latent representations of item $i$ learned from $Net_i$ that can be used for interfering with training process to achieve attack objectives. Similarly, the latent representations $o_{u^*} \in \mathbb{R}^{d_2\times1}$ of user $u^*$ can be learned from $Net_{u^*}$. Furthermore, the dropout is introduced in the two networks to prevent overfitting.

\textbf{\textit{Preference Prediction.}} The latent representations of user and item are then mapped to a shared feature space and  the interaction between user $u^*$ and item $i$ is modelled as: 

\vspace{-1em}
\begin{equation}
o_{u^*i} = \text{concatenate}(o_{u^*}; o_i),
\vspace{-0.5em}
\end{equation} where $o_{u^*i} \in \mathbb{R}^{2d_2\times1}.$ Unlike DeepCoNN using linear FM to predict ratings, we add hidden layers of non-linear transformation between the final prediction layer $f_{l_2}(\cdot)$ and $o_{ui}$ for more accurate preference predictions:

\vspace{-1em}
\begin{equation}
\widehat{r_{u^*i}} = f_{l_2}(\cdots(f_2(f_1(o_{u^*i}))\cdots),
\vspace{-0.5em}
\end{equation} where $\widehat{r_{u^*i}}$ is the predicted rating, $f_j(x)=\text{ReLU}(W_jx+b_j), \ j=1, 2, ..., l_2-1$. As discussed in Section \ref{threat_model}, we focus on top-$K$ recommendation problems, which often require implicit feedback \cite{xu2021understanding}.  To this end, R-Trojan uses \textit{sigmoid} as the activation function of the output layer, as the existing works do (e.g., \cite{huang2021data}). Hence,
$f_{l_2}(x)=\sigma(W_{l_2}x+b_{l_2})$. $W_j$. $b_j$, $W_{l_2}$ and $b_{l_2}$ are trainable parameters.

\subsection{Detection Module} Inspired by the vanilla GAN \cite{goodfellow2020generative}, the detection module plays a minimax game with the generation module empowered by the surrogate module to enhance attack imperceptibility. As shown in Fig. \ref{fig:r_trojan}(c), the module first extracts features from the reviews and ratings of the user profile using TextCNN 
and Encoder, 
respectively, and then fuses these features as the normal/fake user representation. In the case of the fake user $\widetilde{u}$, his/her pattern $h_{\widetilde{u}} \in \mathbb{R}^{2d_3 \times 1}$ is represented as

\vspace{-1.5em}
\begin{equation}
h_{\widetilde{u}} = \text{concatenate}(\text{TextCNN}_{\phi_1}(\delta_{\widetilde{u}}); E_{\phi_2}(r_{\widetilde{u}})),
\vspace{-0.4em}
\end{equation} where $\delta_{\widetilde{u}}$ is $\widetilde{u}$'s review document that contains all fake reviews written by $\widetilde{u}$ and $r_{\widetilde{u}} \in \mathbb{R}^{n}$ denotes fake ratings of $\widetilde{u}$'s profile. Similarly, the normal user $u$'s pattern can be represented as $h_{u} \in \mathbb{R}^{2d_3 \times 1}$. The module finally adopts a MLP to distinguish fake profiles from normal profiles as much as possible:

\vspace{-1em}
\begin{equation}
\widehat{r_{\widetilde{u}}} = f_{l_3}(\cdots(f_2(f_1(h_{\widetilde{u}}))\cdots),
\vspace{-0.4em}
\end{equation} where $\widehat{r_{\widetilde{u}}}$ is the prediction of user $\widetilde{u}$'s profile, $f_j(x)=\text{ReLU}(W_jx+b_j), \ j=1, 2, ..., l_3-1$. Since this module focuses on binary classification tasks, i.e., whether the profile is from a real user or a fake user, the final layer $f_{l_3}(x)=\sigma(W_{l_3}x+b_{l_3})$. $W_j$, $b_j$, $W_{l_3}$ and $b_{l_3}$ are learnable parameters.

\subsection{Learning}

As discussed in Section \ref{threat_model}, we obtain high-quality attack profiles $\widetilde{M_{\widetilde{u}}}$ by solving a bi-level optimization problem.

\textit{\textbf{The Lower-level Problems.}} The original loss in review-based RSs is mean squared error (MSE), which is not suitable for an implicit top-$K$ recommendation task. We adopt binary cross-entropy (BCE) \cite{he2017neural}  for $L_{RS}$, as the existing works do \cite{huang2021data}.  $\Theta$ will be obtained when $L_{RS}$ is minimum that is closest to the actual situation. Similarly, we adopt BCE \cite{lin2022shilling} for $L_{DE}$, due to the detection module is a binary classification task. Since the module plays minimax game with the generation module, $\Phi$ will be obtained when $L_{DE}$ is maximum, indicating that the attack profiles are most similar to normal profiles.



\textit{\textbf{The Upper-level Problem.}} $L_{tran}$ aims to maximize $h(t)$. If $t$ is in the recommendation lists of the normal users (denoted as $\widehat{I_u}$), it is not necessary to optimize much. But if not, it is to minimize the prediction rating gap between $t$ and the items that are in the recommended lists, so that the target item can be promoted to as many normal users as possible. 

\vspace{-1.2em}
\begin{equation} L_{trans}  = 
\left\{
    \begin{array}{ll}
        \log(\sum\limits_{u \in U_t} \sum\limits_{i \in \widehat{I_u}} (e^{\widehat{r_{ui}}} - e^{\widehat{r_{ut}}})+1), & \text{if} \ t \notin  \widehat{I_u}; \\
        C, & \text{if} \ t \in \widehat{I_u},
    \end{array}
\right.
\vspace{-0.3em}
\end{equation} where $U_t = \{ u \in m: r_{ut} = 0 \}$ that is the target user set, $\widehat{r_{ui}}$ is predicted from $RS_\Theta$, $C$ is a small constant, $e^{(\cdot)}$ is used to amplify the rating gap and $\log(\cdot)$ is used to shrink the overall sum of the gap to a range to avoid this attack objective overly dominating the optimization direction. To make sure the loss is positive, we add 1 to the sum. $L_{imper}$ is optimized by tricking the detection module that fake user profiles are from real users.

\begin{table}
  \caption{Statistics of Datasets}
  \vspace{-1.4em}
  \begin{center}
  \small
  \label{tab:statistics}
  \resizebox{.8\columnwidth}{!}{
 \begin{tabular}{c|ccc|c}
\hline
  Dataset & \#Users & \#Items & \#Reviews & Sparsity \\
\hline
  Amazon Musical Instruments & 1,429 & 900 & 10,261 & 99.20\% \\ 
 Amazon Automotive & 2,928 & 1,835 & 20,473 & 99.62\% \\ 
  Yelp & 1,599 & 1,318 & 30,120 & 98.57\%\\ 
\hline
  \end{tabular}}
  \end{center}
   \vspace{-2.6em}
\end{table}

\begin{table*}
\caption{HR@10 and NDCG@10 of different attacks against various victim RSs on real-world datasets (attack size 3\%). 
}
  \vspace{-1.2em}
  \begin{center}
  \small
  \label{tab:comp_results}
  \resizebox{1.7\columnwidth}{!}{
  \begin{tabular}{c|c|c|c|c|c|c|c|c|c|c|c|c}
    \hline
     \multirow{2}{*}{Victim RS}  & \multirow{2}{*}{Dataset} & \multirow{2}{*}{Metric} & \multicolumn{10}{c}{Attack Method}\\

     \cline{4-13}
     
     & & &  {Random} & {Bandwagon}& {PGA}& {DCGAN} & {AUSH} & {DLA} & {RecUP} &  {Leg-UP}& {TrialAttack} &{R-Trojan}\\

    \hline
    \hline
    
     \multirow{6}{*}{WRMF} & \multirow{2}{*}{Musical} & HR  & 0.2649 & 0.2968 & 0.3904 & 0.3599 & 0.3781 & 0.4383 &  0.4020 & 0.4681 &  \underline{0.4804} & \textbf{0.5530}\\
     
     & &  NDCG  & 0.1082 & 0.1211 & 0.1822 &  0.1957 & 0.1845 & 0.2208 &  0.1970 & 0.2374 & \underline{0.2418} & \textbf{0.3617}\\

     \cline{2-13}

     & \multirow{2}{*}{Automotive} & \multicolumn{1}{c|}{HR} & 0.1137 & \multicolumn{1}{c|}{0.1348} & 0.1563 & \multicolumn{1}{c|}{0.1695} & 0.1698 & \multicolumn{1}{c|}{0.2094} & 0.1730 & \multicolumn{1}{c|}{\underline{0.2305}} & 0.2284  & \textbf{0.2693}\\
     
      & & \multicolumn{1}{c|}{NDCG} & 0.0451 & \multicolumn{1}{c|}{0.0533} & 0.0772 & \multicolumn{1}{c|}{0.0818} & 0.0796 & \multicolumn{1}{c|}{0.0981} & 0.0781 & \multicolumn{1}{c|}{\underline{0.1232}} & 0.1145 & \textbf{0.1677}\\

     \cline{2-13}

     & \multirow{2}{*}{Yelp} & \multicolumn{1}{c|}{HR} & \multicolumn{1}{c|}{0.0939} & 0.0861 & \multicolumn{1}{c|}{0.1082} & 0.1017 & \multicolumn{1}{c|}{0.1231} & 0.1276 & \multicolumn{1}{c|}{0.1399} & 0.1360 & \multicolumn{1}{c|}{\underline{0.1464}} & \textbf{0.1852} \\ 
     
      & & \multicolumn{1}{c|}{NDCG} &  0.0635 & \multicolumn{1}{c|}{0.0563} & 0.0801 & \multicolumn{1}{c|}{0.0731} & 0.0799 & \multicolumn{1}{c|}{0.0893} & 0.0969  & \multicolumn{1}{c|}{\underline{0.1087}} & 0.0998 & \textbf{0.1496} \\

    \hline

      \multirow{6}{*}{NCF} & \multirow{2}{*}{Musical} & \multicolumn{1}{c|}{HR} & 0.1495 & \multicolumn{1}{c|}{0.1509} & 0.1633 & \multicolumn{1}{c|}{0.1807} & 0.2068 & \multicolumn{1}{c|}{0.2525} & 0.2104 & \multicolumn{1}{c|}{0.2438} & \underline{0.2714} & \textbf{0.3382}\\

     & & \multicolumn{1}{c|}{NDCG}   & 0.0633 & \multicolumn{1}{c|}{0.0575} & 0.0677 & \multicolumn{1}{c|}{0.0816} & 0.0944 & \multicolumn{1}{c|}{0.1305} & 0.0802 & \multicolumn{1}{c|}{0.1112} & \underline{0.1503} & \textbf{0.2009}\\

     \cline{2-13}

     & \multirow{2}{*}{Automotive} & \multicolumn{1}{c|}{HR}  & 0.1043 & \multicolumn{1}{c|}{0.1102} & 0.1192 & \multicolumn{1}{c|}{0.1383} & 0.1650 & \multicolumn{1}{c|}{0.1962} & 0.1827 & \multicolumn{1}{c|}{0.2038} & \underline{0.2187} & \textbf{0.2679} \\ 
     
      & & \multicolumn{1}{c|}{NDCG} & 0.0474 & \multicolumn{1}{c|}{0.0515} & 0.0604 & \multicolumn{1}{c|}{0.0854} & 0.0971 & \multicolumn{1}{c|}{0.1009} & 0.0935 & \multicolumn{1}{c|}{0.1109} & \underline{0.1353} & \textbf{0.1594}\\

     \cline{2-13}

     & \multirow{2}{*}{Yelp} & \multicolumn{1}{c|}{HR}  & 0.0784 & \multicolumn{1}{c|}{0.0667} & 0.0848 & \multicolumn{1}{c|}{0.0965} &  0.0991 & \multicolumn{1}{c|}{0.1159} & 0.1082 & \underline{0.1328} & 0.1250 & \textbf{0.1729}\\ 
     
      &  & \multicolumn{1}{c|}{NDCG}  & 0.0472 & \multicolumn{1}{c|}{0.0337} & 0.0561 & \multicolumn{1}{c|}{0.0603} & 0.0553 & \multicolumn{1}{c|}{0.0752} & 0.0598 & 0.0844 & \underline{0.0851} & \textbf{0.1162}\\

      \hline

  \multirow{6}{*}{LightGCN} & \multirow{2}{*}{Musical} & \multicolumn{1}{c|}{HR} & \multicolumn{1}{c|}{0.0602} & 0.0718 & \multicolumn{1}{c|}{0.0893} & 0.0972 & \multicolumn{1}{c|}{0.1110} & \underline{0.1509} & \multicolumn{1}{c|}{0.1023} & 0.1437 & \multicolumn{1}{c|}{0.1357} & \textbf{0.1858}\\

     & & \multicolumn{1}{c|}{NDCG}  & \multicolumn{1}{c|}{0.0260} & 0.0370 & \multicolumn{1}{c|}{0.0413} & 0.0466 & \multicolumn{1}{c|}{0.0556} & \underline{0.0786} & \multicolumn{1}{c|}{0.0517} & 0.0698 & \multicolumn{1}{c|}{0.0732} & \textbf{0.0983} \\

     \cline{2-13} 
     
     & \multirow{2}{*}{Automotive} & \multicolumn{1}{c|}{HR} & \multicolumn{1}{c|}{0.0340} & 0.0343 & \multicolumn{1}{c|}{0.0589} & 0.0433 & \multicolumn{1}{c|}{0.0634} & \underline{0.1102} & \multicolumn{1}{c|}{0.0735} & 0.1088 & \multicolumn{1}{c|}{0.1009} & \textbf{0.1536} \\ 
     
      & & \multicolumn{1}{c|}{NDCG} & \multicolumn{1}{c|}{0.0189} & 0.0174  & \multicolumn{1}{c|}{0.0359} & 0.0238 & \multicolumn{1}{c|}{0.0324} & 0.0599 & \multicolumn{1}{c|}{0.0387} & 0.0632 & \multicolumn{1}{c|}{\underline{0.0681}} & \textbf{0.0846} \\

     \cline{2-13}

     & \multirow{2}{*}{Yelp} & \multicolumn{1}{c|}{HR} & \multicolumn{1}{c|}{0.0136} & 0.0117 & \multicolumn{1}{c|}{0.0246} &  0.0317 & \multicolumn{1}{c|}{0.0415} & \underline{0.0602} & \multicolumn{1}{c|}{0.0434} & 0.0544 & \multicolumn{1}{c|}{0.0512} &  \textbf{0.0926} \\ 
     
      & & \multicolumn{1}{c|}{NDCG} & \multicolumn{1}{c|}{0.0059} & 0.0061 & \multicolumn{1}{c|}{0.0141} & 0.0197 & \multicolumn{1}{c|}{0.0265} & 0.0262  & \multicolumn{1}{c|}{0.0252} &  0.0354 & \multicolumn{1}{c|}{\underline{0.0359}} &  \textbf{0.0560} \\ 

      \hline
  \end{tabular}}
  \end{center}
 \vspace{-3.1em}
\end{table*}

\vspace{-0.8em}
\begin{equation}
    L_{imper} = \frac{1}{\widetilde{m}} \sum\limits_{\widetilde{u} \in \widetilde{m}} \log(1-DE_{\Phi}(\widetilde{M_{\widetilde{u}}})),
\vspace{-0.3em}
\end{equation} where the loss is weighted by $\lambda$ and combined with $L_{trans}$ to form the final loss for optimizing the parameters of the generation module, as detailed in Section \ref{threat_model}. Since \cite{lin2022shilling} also uses a surrogate RS, our training procedure follows this work. Adam optimization algorithm is used in our implementation.

\section{Experiments} \label{experiments}

\subsection{Experimental Setup}

\subsubsection{Datasets Selection}
We use three real-world datasets from different scenarios  \cite{mcauley2015image} to evaluate R-Trojan, as detailed in Table \ref{tab:statistics}. For Yelp\footnote{https://www.yelp.com/dataset}, we randomly select a subset to avoid exceeding the hardware limit.
We adopt the leave-one-out method \cite{he2017neural} to select the test set, and the rest is regarded as the training set and the validation set, with a ratio of 9:1.

\subsubsection{Baseline Attack Methods}
We compare R-Trojan with typical and state-of-the-art poisoning attacks including Random, Bandwagon \cite{gunes2014shilling}, PGA \cite{li2016data}, DCGAN \cite{christakopoulou2019adversarial}, AUSH \cite{lin2020attacking},  DLA \cite{huang2021data}, RecUP \cite{zhang2021attacking}, Leg-UP \cite{lin2022shilling} and TrialAttack \cite{wu2021triple}.

\subsubsection{Targeted Recommender Systems}
We consider three rating-only-based victim RSs, as existing works do \cite{lin2022shilling, lin2020attacking}: WRMF \cite{hu2008collaborative}, NCF \cite{he2017neural} and state-of-the-art LightGCN \cite{he2020lightgcn}. Moreover, to evaluate the attack effectiveness on review-based RSs under both black-box and white-box settings, we use DeepCoNN \cite{zheng2017joint} and our surrogate model as victim RSs.

\subsubsection{Evaluation Metrics}
Following the existing works \cite{zhang2021attacking, lin2020attacking, huang2021data, lin2022shilling, wu2021triple}, we use two ranking metrics to evaluate attack effectiveness: hit ratio (HR@$k$) and normalized discounted cumulative gain (NDCG@$k$). $k$ is set to 10 here.

\subsubsection{Data Pre-processing and Hyper-Parameter Settings}
We adopt some data pre-processing strategies to enhance the effectiveness of evaluation. To better catch the semantic information from reviews, each review is encoded by pre-trained 300-dimensional word embeddings\footnote{https://code.google.com/archive/p/word2vec/}.
Moreover, we binarize ratings of the attack profiles before feeding them to the RS so that the profiles are suitable for top-N implicit recommendation tasks \cite{xu2021understanding, huang2021data}.  Note that we still maintain the original explicit ratings when evaluating the imperceptibility.

To maintain a fair comparison, we set $A$ to 3\% of the population as default \cite{lin2022shilling} 
and $F$ equals to the average number of ratings per user in the data set for all attacks. The other hyper-parameters of baselines are optimized by grid search on the validation set. The hyper-parameters of the victim RS and GPT-2 are set as suggested in the original papers and Hugging Face\footnote{https://github.com/huggingface/transformers} respectively. For R-Trojan, we set $l_1$=$l_2$=3, the number of layers 
in the detection module to 2, $d_2$=$d_3$=50, training epochs to 20, batch size to 256, learning rate to 0.001, $\lambda$ to 0.5 to balance the two attack goals, dropout to 0.5, and randomly sample 4 negative instances per positive instance. Moreover, the parameters of TextCNNs are inherited from DeepCoNN. 
\vspace{-0.5em}

\begin{figure}
\centering
\begin{subfigure}{.24\textwidth}
    \centering
    \includegraphics[width=\linewidth]{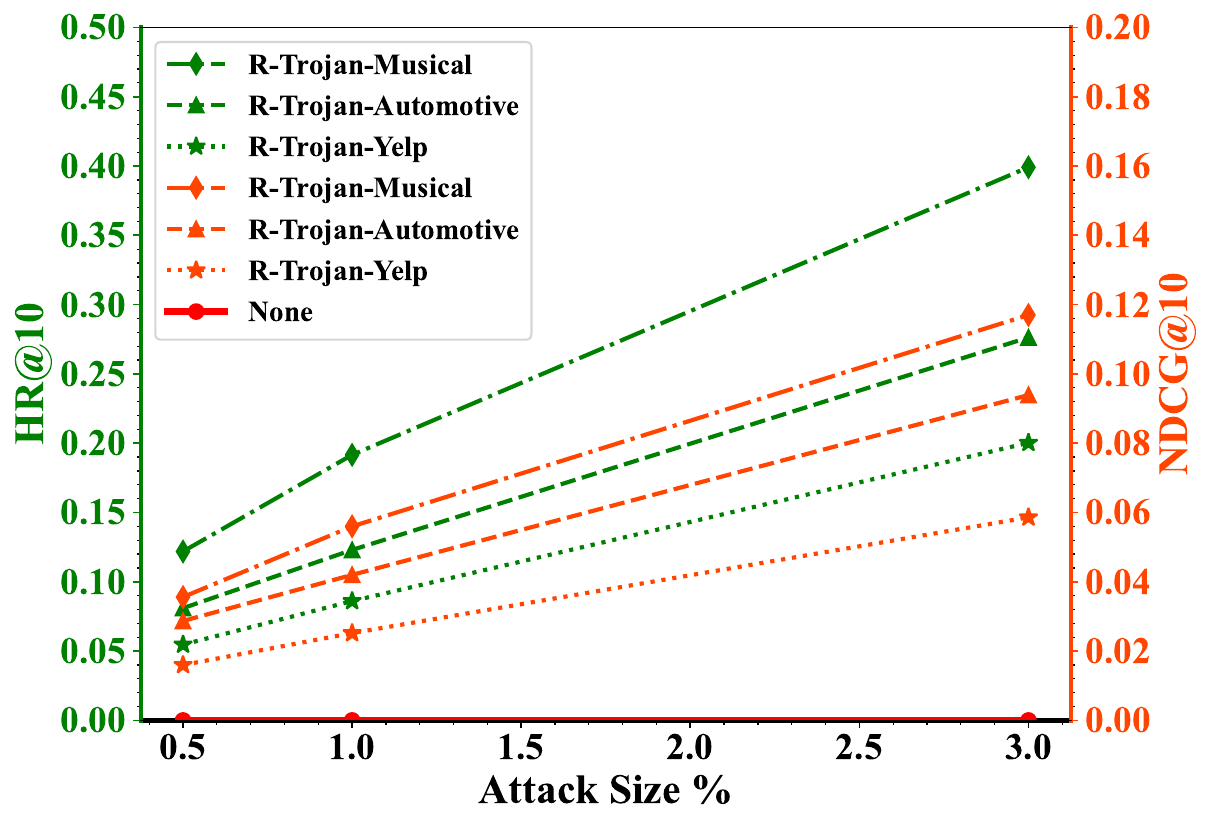} 
    \vspace{-1.5em}
    \caption{R-Trojan on DeepCoNN}
    \label{SUBFIGURE LABEL 3}
\end{subfigure}
\begin{subfigure}{.24\textwidth}
    \centering
    \includegraphics[width=\linewidth]{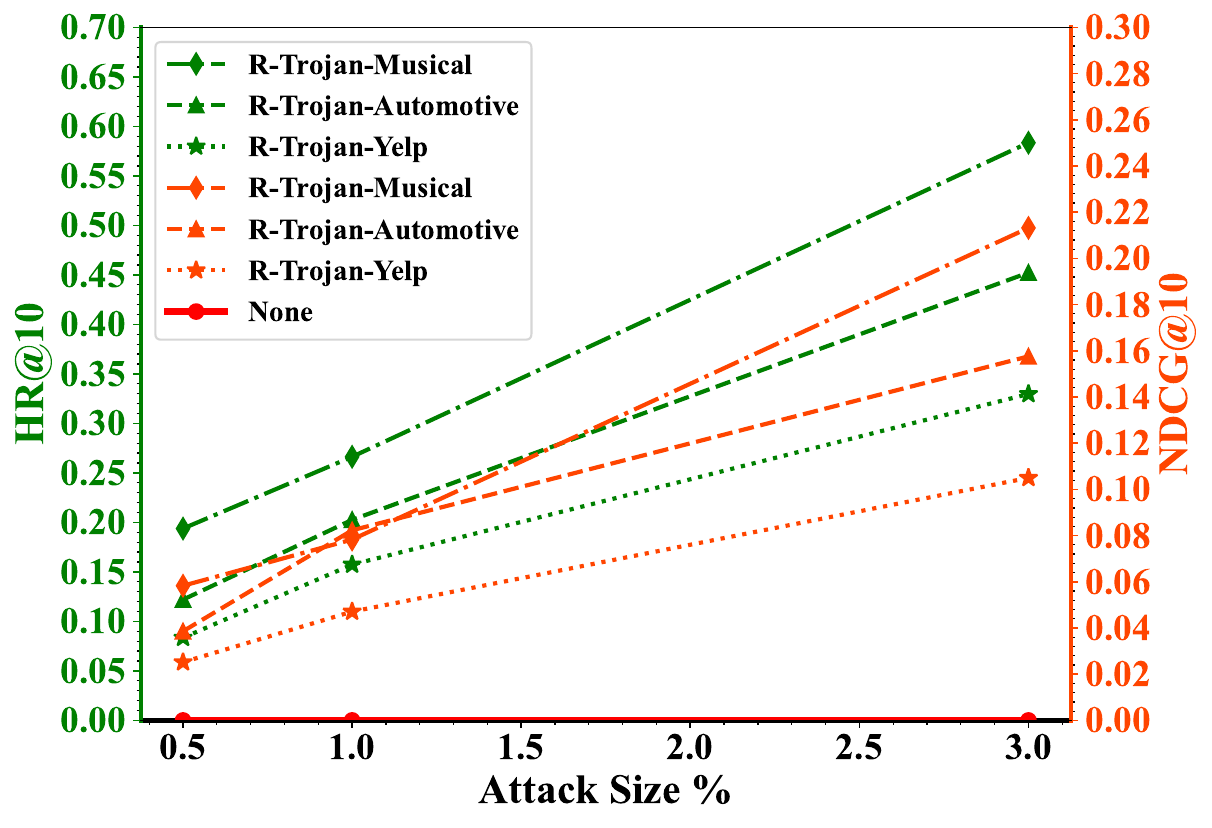} 
    \vspace{-1.5em}
    \caption{R-Trojan on DeepCoNN++}
    \label{SUBFIGURE LABEL 4}
\end{subfigure}
\vspace{-1.6em}
\caption{R-Trojan performance with different attack sizes against various review-based RSs on three real-world datasets.}
\label{fig:attack_review_RS}
\vspace{-1.8em}
\end{figure}

\subsection{Attack Transferability}

Table \ref{tab:comp_results} and Fig. \ref{fig:attack_review_RS} demonstrate R-Trojan excellent transferability, which shows that reviews can effectively enhance attack profiles. R-Trojan greatly outperforms the state-of-the-art baselines against different victim RSs on real-world datasets by achieving higher HR@10 and NDCG@10, as illustrated in Table \ref{tab:comp_results}. Leg-UP, TrialAttack and DLA show comparable performance. The reasons may be that both Leg-UP and TrialAttack adopt a surrogate RS, which helps them improve transferability, and DLA is an attack method optimized for DL-based RSs. RecUP and AUSH perform worse than them, but better than DCGAN, due to they are tailored GANs for shilling attacks. Since PGA is designed for MF, it cannot effectively transfer to DL-based RSs such as NCF and LightGCN, as shown in the table. Among baselines, conventional attacks (i.e., Random and Bandwagon) are the least transferable. 

\subsection{Poisoning Review-based RSs}

As shown in Fig. \ref{fig:attack_review_RS}, the effectiveness of R-Trojan on DeepCoNN and our surrogate RS, compared to None (i.e., the circumstance when no attack is launched), demonstrates that review-based RSs are vulnerable to poisoning attacks, especially in white-box settings. We also set $A = 0.5\%$ and $1\%$. With the increase in attack size leads to the increase in the popularity of $t$, $t$ is promoted to more normal users.

\subsection{Attack Imperceptibility}

\begin{figure}
\centering
\begin{subfigure}{.24\textwidth}
    \centering
    \includegraphics[width=0.65\linewidth]{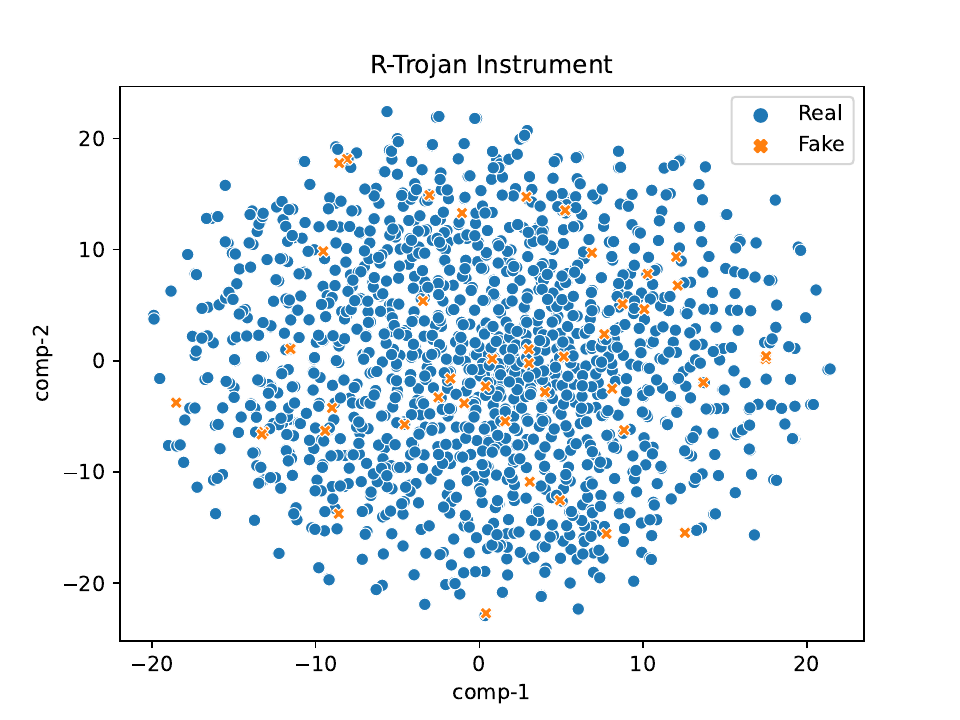}  
    \vspace{-0.4em}
    \caption{R-Trojan on Musical}
    \label{SUBFIGURE LABEL 1}
\end{subfigure}
\begin{subfigure}{.24\textwidth}
    \centering
    \includegraphics[width=0.65\linewidth]{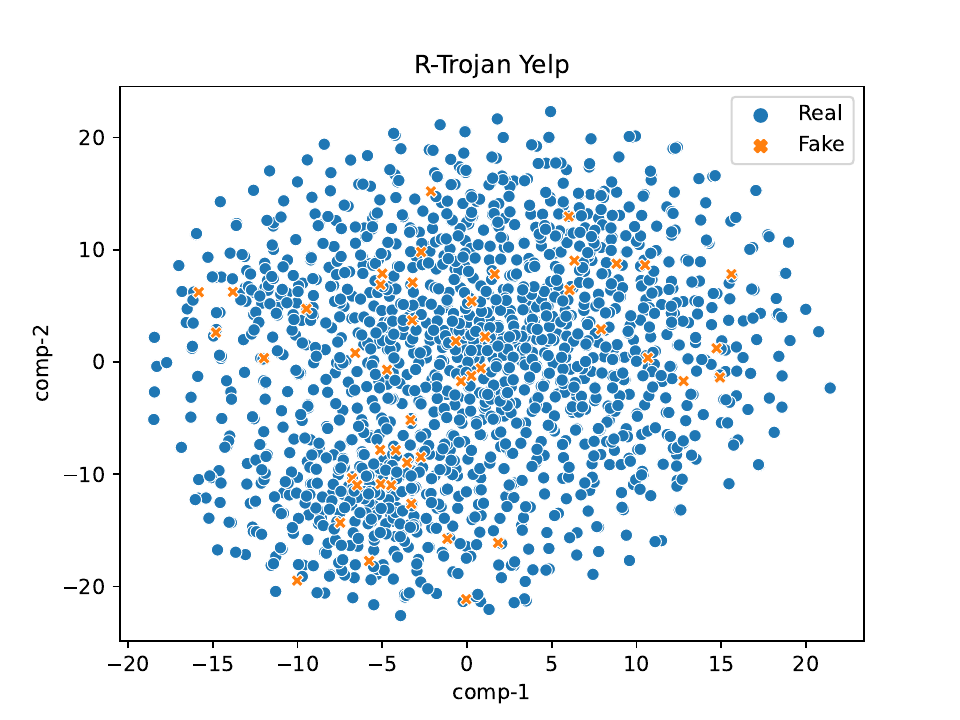}
    \vspace{-0.4em}
    \caption{R-Trojan on Yelp}
    \label{SUBFIGURE LABEL 2}
\end{subfigure}
\vspace{-1.5em}
\caption{Visualization of R-Trojan's fake user profiles and real user profiles on real-world datasets.}
\label{fig:tsne_graphs}
\vspace{-1.8em}
\end{figure}

We use t-SNE \cite{lin2022shilling} to visualize real user's and R-Trojan's profile representations generated by the detection module
on real-world datasets, as shown in Fig. \ref{fig:tsne_graphs}. The fake user profiles are scattered in the area of real user profiles, which indicates that fake profiles have a similar distribution as real profiles. As a result, R-Trojan can launch invisible attacks.

\section{Conclusion}

We propose a novel review-enhanced poisoning attack  framework R-Trojan  in this paper, which leverages transformers for an improved GAN, for generating high-quality fake user profiles. 
R-Trojan clearly outperforms the state-of-the-art GAN-based attack methods on real-world datasets for various RSs. It also has good imperceptibility. Moreover, we provide a pioneer study on poisoning attacks to review-based RSs.

\renewcommand{\baselinestretch}{0.9}

\bibliographystyle{IEEEtran}
\bibliography{IEEEabrv,IEEEexample}

\begin{thebibliography}{10}
\providecommand{\url}[1]{#1}
\csname url@samestyle\endcsname
\providecommand{\newblock}{\relax}
\providecommand{\bibinfo}[2]{#2}
\providecommand{\BIBentrySTDinterwordspacing}{\spaceskip=0pt\relax}
\providecommand{\BIBentryALTinterwordstretchfactor}{4}
\providecommand{\BIBentryALTinterwordspacing}{\spaceskip=\fontdimen2\font plus
\BIBentryALTinterwordstretchfactor\fontdimen3\font minus
  \fontdimen4\font\relax}
\providecommand{\BIBforeignlanguage}[2]{{%
\expandafter\ifx\csname l@#1\endcsname\relax
\typeout{** WARNING: IEEEtran.bst: No hyphenation pattern has been}%
\typeout{** loaded for the language `#1'. Using the pattern for}%
\typeout{** the default language instead.}%
\else
\language=\csname l@#1\endcsname
\fi
#2}}
\providecommand{\BIBdecl}{\relax}
\BIBdecl

\bibitem{gunes2014shilling}
I.~Gunes, C.~Kaleli, A.~Bilge, and H.~Polat, ``Shilling attacks against
  recommender systems: a comprehensive survey,'' \emph{Artificial Intelligence
  Review}, 2014.

\bibitem{li2016data}
B.~Li, Y.~Wang, A.~Singh, and Y.~Vorobeychik, ``Data poisoning attacks on
  factorization-based collaborative filtering,'' \emph{NIPS}, 2016.

\bibitem{fang2018poisoning}
M.~Fang, G.~Yang, N.~Z. Gong, and J.~Liu, ``Poisoning attacks to graph-based
  recommender systems,'' in \emph{ACSAC}, 2018.

\bibitem{christakopoulou2019adversarial}
K.~Christakopoulou and A.~Banerjee, ``Adversarial attacks on an oblivious
  recommender,'' in \emph{RecSys}, 2019.

\bibitem{zhang2021attacking}
X.~Zhang, J.~Chen, R.~Zhang, C.~Wang, and L.~Liu, ``Attacking recommender
  systems with plausible profile,'' \emph{TIFS}, 2021.

\bibitem{lin2020attacking}
C.~Lin, S.~Chen, H.~Li, Y.~Xiao, L.~Li, and Q.~Yang, ``Attacking recommender
  systems with augmented user profiles,'' in \emph{CIKM}, 2020.

\bibitem{huang2021data}
H.~Huang, J.~Mu, N.~Z. Gong, Q.~Li, B.~Liu, and M.~Xu, ``Data poisoning attacks
  to deep learning based recommender systems,'' in \emph{NDSS}, 2021.

\bibitem{wu2021triple}
C.~Wu, D.~Lian, Y.~Ge, Z.~Zhu, and E.~Chen, ``Triple adversarial learning for
  influence based poisoning attack in recommender systems,'' in \emph{KDD},
  2021.

\bibitem{lin2022shilling}
C.~Lin, S.~Chen, M.~Zeng, S.~Zhang, M.~Gao, and H.~Li, ``Shilling black-box
  recommender systems by learning to generate fake user profiles,''
  \emph{TNNLS}, 2022.

\bibitem{goodfellow2020generative}
I.~Goodfellow, J.~Pouget-Abadie, M.~Mirza, B.~Xu, D.~Warde-Farley, S.~Ozair,
  A.~Courville, and Y.~Bengio, ``Generative adversarial networks,''
  \emph{Communications of the ACM}, 2020.

\bibitem{xu2021understanding}
Z.~Xu, H.~Zeng, and Q.~Ai, ``Understanding the effectiveness of reviews in
  e-commerce top-n recommendation,'' in \emph{ICTIR}, 2021.

\bibitem{mohawesh2021fake}
R.~Mohawesh, S.~Xu, S.~N. Tran, R.~Ollington, M.~Springer, Y.~Jararweh, and
  S.~Maqsood, ``Fake reviews detection: A survey,'' \emph{IEEE Access}, 2021.

\bibitem{zheng2017joint}
L.~Zheng, V.~Noroozi, and P.~S. Yu, ``Joint deep modeling of users and items
  using reviews for recommendation,'' in \emph{WSDM}, 2017.

\bibitem{zhang2019deep}
S.~Zhang, L.~Yao, A.~Sun, and Y.~Tay, ``Deep learning based recommender system:
  A survey and new perspectives,'' \emph{CSUR}, 2019.

\bibitem{radford2019language}
A.~Radford, J.~Wu, R.~Child, D.~Luan, D.~Amodei, I.~Sutskever \emph{et~al.},
  ``Language models are unsupervised multitask learners,'' \emph{OpenAI}, 2019.

\bibitem{he2017neural}
X.~He, L.~Liao, H.~Zhang, L.~Nie, X.~Hu, and T.-S. Chua, ``Neural collaborative
  filtering,'' in \emph{WWW}, 2017.

\bibitem{mcauley2015image}
J.~McAuley, C.~Targett, Q.~Shi, and A.~Van Den~Hengel, ``Image-based
  recommendations on styles and substitutes,'' in \emph{SIGIR}, 2015.

\bibitem{hu2008collaborative}
Y.~Hu, Y.~Koren, and C.~Volinsky, ``Collaborative filtering for implicit
  feedback datasets,'' in \emph{ICDM}, 2008.

\bibitem{he2020lightgcn}
X.~He, K.~Deng, X.~Wang, Y.~Li, Y.~Zhang, and M.~Wang, ``Lightgcn: Simplifying
  and powering graph convolution network for recommendation,'' in \emph{SIGIR},
  2020.

\end{thebibliography}

\end{document}